# Applying the Spacecraft with a Solar Sail to Form the Climate on a Mars Base

By Miroslav A. ROZHKOV, Irina V. GORBUNOVA and Olga L. STARINOVA [1]

[1] *Institute of Space Rocket Engineering, Samara University, Samara, Russia*

This article is devoted to research the application of the spacecraft with a solar sail to support the certain climatic conditions in an area of the Mars surface. Authors propose principles of functioning of the spacecraft, intended to create a light and thermal light spot in a predetermined area of the Martian surface. The mathematical motion model in such condition of the solar sail's orientation is considered and used for motion simulation session. Moreover, the analysis of this motion is performed. Thus, were obtained parameters of the stationary orbit of the spacecraft with a solar sail and were given recommendations for further applying spacecrafts to reflect the sunlight on a planet's surface.

**Key Words:**   Solar Sail, Motion Simulation, Control Law, Mars, Cylindrical Orbits

## Nomenclature

| | | |
|---|---|---|
| *x, y, z* | : | position in inertial coordinate system |
| K | : | spacecraft's center of mass |
| B | : | current illuminated area |
| **T** | : | thrust |
| **S** | : | position vector of the Sun |
| **a** | : | acceleration |
| **n**$^0$ | : | sail normal unit vector |
| **r** | : | position vector of the solar sail |
| *t* | : | current time |
| Subscripts | | |
| g | : | gravitational |
| *x, y, z* | : | projections on corresponding axes |

## 1. Introduction

The spacecraft with a solar sail (SCSS), regardless of the low thrust, is able to perform a wide field of specific missions. Such as space exploration, setting up and keeping a spacecraft on light-levitated stationary cylindrical orbits, retransmission of energy, removing space debris and illuminating of areas of a planet.

Constant reflection of beam of light off a solar sail on planet's surface haven't been considered as full year mission before. However, Russian Federal Space Agency carried out similar project "Znamya 2", which is better known as "Space Mirror". In 1993, the spaceship 'Progress M-15' placed into orbit a 20-meter film mirror, which deployed and produced a light spot on the blackout side of the Earth. [1]

Authors propose to form the climate on a Mars base by reflecting sunlight off a solar sail on a predetermined area of the surface with the object of producing additional heat and light. Mars was chosen as the most suitable planet for colonization, but it lacks for sunlight due to long distance from the Sun.

## 2. Mathematical simulation

In this paper we describe orientation of the solar sail and its alteration during the motion of the spacecraft, with regard to rotation of the Mars around its spin axis and around the Sun. An ideal solar sail and a spherically symmetric Mars are assumed. Influences from natural satellites Phobos and Deimos are neglected. Consider Mars-centered inertial coordinate system with origin 'O' at the Mars' center of mass as shown in Fig. 1. The system is an inertial frame with axes *x* and *y* in the equatorial plane and the *z* axis is directed along the Mars' spin axis. Furthermore, the *x* axis is aligned with the vernal equinox.

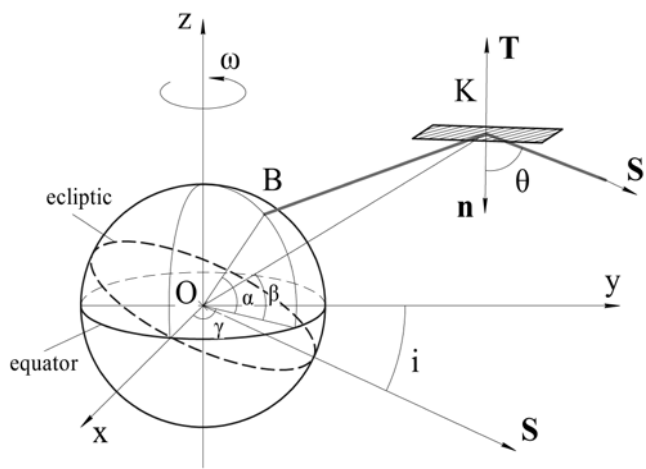

Fig. 1.   The considered inertial coordinate system.

Latitude and longitude are considered as angles α and γ and determine the position of the illuminated area on the Mars surface. Meanwhile, the SCSS is located on the stationary orbit of Mars, which is called "areostationary orbit" (ASO) and has radius *R* = 20399 km. The ASO is necessary condition for constant illuminating an established area, but it's located in the equatorial plane, thereby for areas with high latitude this spacecraft's position increases the incidence angle of reflected light produced by a solar sail. Therefore is suggested to use levitated cylindrical orbits, which will arise due to vertical (along the $z$ axis) component of the thrust.[2-4] These non-Keplerian orbits have their own low latitude (angle β)

Position of the Sun and its motion in Mars-centered inertial coordinate system is defined by equations of transformation from osculating orbit into inertial frame. The longitude of ascending node is equal zero, thus a position of the Sun can be written in component form as

$$\begin{cases} S_x = |\mathbf{S}| \cdot \cos(u), \\ S_y = |\mathbf{S}| \cdot \sin(u) \cdot \cos(i), \\ S_z = |\mathbf{S}| \cdot \sin(u) \cdot \sin(i). \end{cases} \quad (1)$$

Axial tilt of Mars is $i = 25°$ and angle of the true anomaly $u$ is determined by

$$u = \omega_s \cdot t, \quad (2)$$

where $\omega_s = 2\pi/686.96 \cdot 86400$ rad/s is angular velocity of the Sun with respect to the inertial frame.

Sail normal vector is bisector of angle included between incident and reflected rays of light in condition that reflected rays create light spot on certain area of the Mars surface. The bisector of angle included between two vectors is defined as

$$\mathbf{n} = \frac{\mathbf{KB}}{|\mathbf{KB}|} + \frac{\mathbf{KS}}{|\mathbf{KS}|}. \quad (3)$$

Orientation control of the solar sail is possible due to ability of sail area elements to change surface reflection characteristics of sunlight,[5] consequently create a force moment to operate a solar sail plane.

## 3. Motion in condition of reflection of the sunlight

### 3.1. Equation of motion

Equation of motion in the Mars-centered inertial coordinate system is given by

$$\frac{d^2\mathbf{r}}{dt^2} + \mathbf{a}_g = \mathbf{a}. \quad (4)$$

The acceleration produced by sail from solar radiation pressure is defined as

$$\mathbf{a} = a_0 \cos^2(\theta) \cdot \mathbf{n}^0. \quad (5)$$

where $a_0$ is the sail characteristic acceleration generated by SCSS during direct orientation toward the Sun, and $\theta$ is the incidence angle of the sunlight on a solar sail.

For this work we considered ideal solar sails with characteristic acceleration $a_0 = 8$ mms$^{-2}$. Such acceleration would be reachable thanks to monoatomic layered material, which is graphene.[6] The areal mass (the ratio of the mass of the spacecraft to the area of the sail) of a graphene-based sail would be about 0.02 g/m$^2$.[7]

We simulate movement of the SCSS, which performs illumination of the area with zero coordinates from ASO. Also we take initial location of the Sun close to winter solstice. Modeling solution performed numerically in the inertial coordinate system by using classical Runge-Kutta method.

Solution of the Eq. (4) shows, that the component of the sail acceleration parallel to the equatorial plane $a_p$ changes orbit shape of the ASO toward elliptic. This orbit, which is shown in Fig. 2, appeared in consequence of acceleration along the speed direction of the SCSS at the periapsis point, thus increases radius of the apoapsis. On the other hand this parallel component of acceleration at the apoapsis point directs against the speed of the spacecraft and decreases the radius of periapsis.

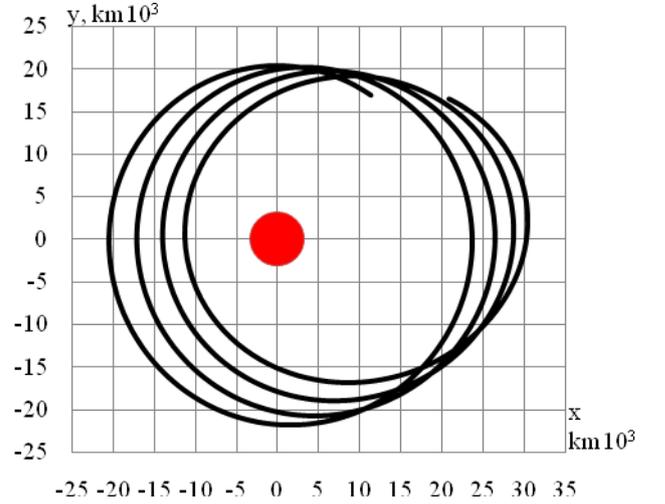

Fig. 2. Change of the orbit shape due to influence of the parallel component of acceleration $a_p$.

### 3.2. Motion without the parallel component of sail acceleration

Transversal (projection on the tangent to the trajectory) component of acceleration provides significant influence on the alteration of orbit shape. However, it's impossible to orientate the solar sail both without producing transversal component of sail acceleration and with reflection of the sunlight toward to the established area on the Mars surface.

To accomplish assigned task authors propose to use additional propeller with low thrust and ability for continuously work during one year or at least half-year mission. This propeller will be able to offset the disturbing parallel component of acceleration. It can be one or few hall effect thrusters (also called stationary plasma thrusters SPT), such as SPT-290 or less powerful analog. This engine's thrust can reach 1.5 N, with its mass around 23 kg and operational life 27000 hours.[8] Such plasma propulsion engine located at the solar sail's center of mass, with ability to deflect on 25°, will be able to keep form of the cylindrical orbit.

Again we simulate movement of the SCSS, with the same initial parameters and 4 SPT-290 working against parallel component of solar sail acceleration.

Because of the asymmetry of the vertical component of acceleration with respect to $z$ axis the SCSS, instead of maintains cylindrical orbit, increases the tilt angle (Fig. 3). This significant change of the acceleration is connected with huge variation in angle $\theta$ during the day, as it's shown in Fig. 4. For example, during two opposite phase of flight, when SCSS creates direct line with the Sun and Mars, angle $\theta$ changes from 78° to 12°. It dramatically influences on acceleration value in accordance with Eq. (5). When orbit of the SCSS has some tilt angle, the vertical component of sail acceleration produces projection on the tangent to the trajectory, thereby the orbit shape begins to change. In Fig. 5 shown such alteration of the orbit shape. In addition, if altitude of the spacecraft will be too high, then the gravitational acceleration $|\mathbf{a}_g|$ can become low enough, compared to the acceleration from solar radiation pressure $|\mathbf{a}|$, that the solar sail will be able to fly away from the Mars' sphere of influence.

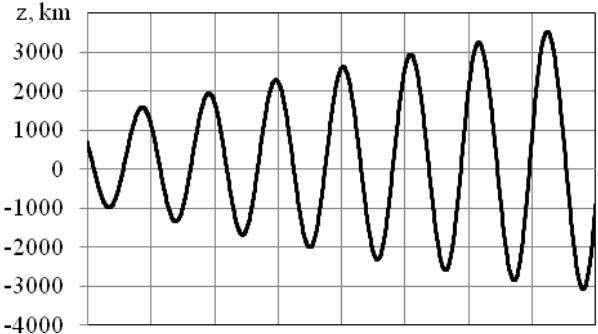

Fig. 3. Tilt angle of the SCSS orbit due to $a_z$, which is shown as variations in altitude from +z to -z.

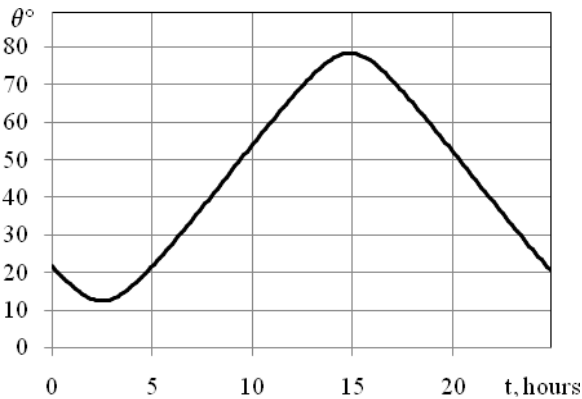

Fig. 4. Change of the angle $\theta$ during one Martian day.

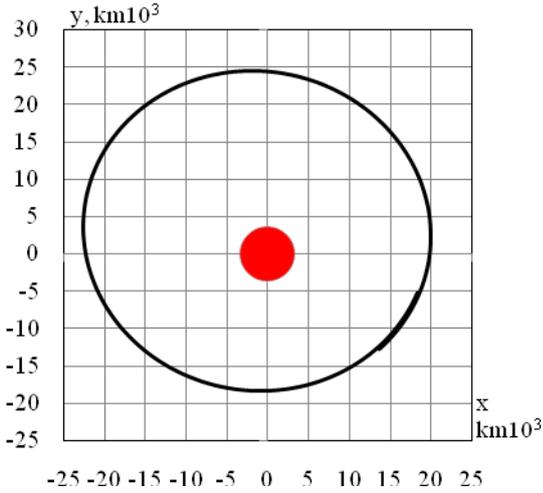

Fig. 5. Change of the areostationary orbit shape.

### 3.3. Equalization of the vertical component of sail acceleration

For maintaining cylindrical orbit it's necessary to have steady value of the vertical component of acceleration produced by solar sail, but that is not reachable with continuous alteration of the angle $\theta$ in consequence of reflection of the sunlight on certain area on Mars surface. Therefore authors propose to use so called "pseudo cylindrical orbit". This orbit appears to be similar to non-Keplerian orbit, but with small variations in altitude of the SCSS above the equatorial plane. To achieve this orbit there is a need to equalize accelerations in symmetrical positions with respect to z axis. It can be achieved if during first 12.31 hours illuminate surface of Mars, and during the other time orientate the solar sail in the same way as in opposite position relative to the line normal to projection of position vector **S** on plane $xOy$ of inertial Mars-centered coordinate system (Fig. 6). Thus, we suggest to use the spacecraft with a solar sail for illuminating the Mars base at night time, and then equalize the differences in vertical component of acceleration all remaining time. This new control law will prevent formation of tilt angle and will keep spacecraft on stationary orbit.

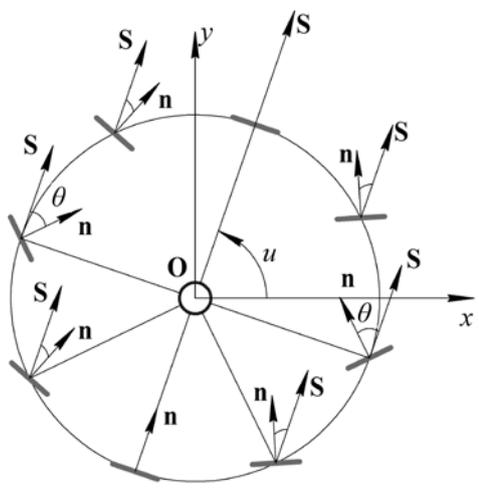

Fig. 6. New control law for the solar sail's orientation.

We simulate movement of the SCSS, with the same initial parameters as before, but with new control low for equalization the differences in vertical component of acceleration. With similar orientation during both night and day time angle $\theta$ has two maximum and minimum values as shown in Fig. 7. Disturbing tilt axis reduced enough, that transversal component of acceleration doesn't change the orbit shape during all year mission.

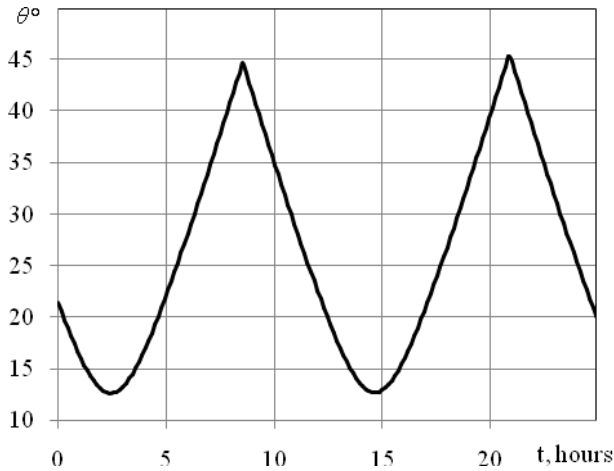

Fig. 7. Alteration of the angle $\theta$ with new control law.

The Graph in Fig. 8 shows significant differences in motion with and without equalization. With new control law were decreased time of illumination during the Martian day, but total time of the mission were increased.

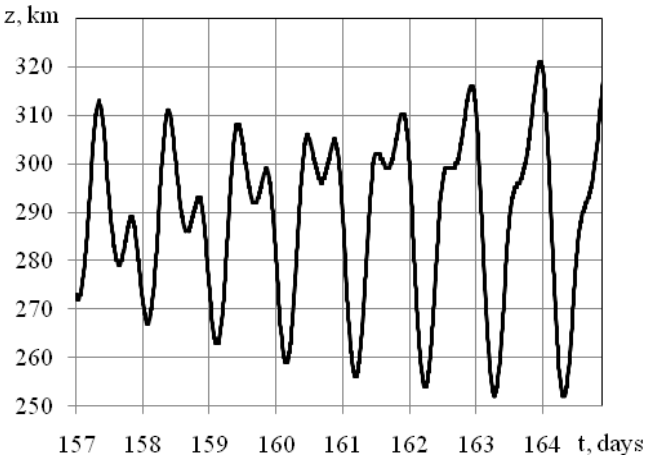

Fig. 8. Change of the SCSS altitude with new control law.

## 4. Simulation analysis

### 4.1. Half-yearly motion

The solar sail motion with established latitude of the Mars base requires certain initial time to start the illumination of planet's surface. This time is defined as a moment for two opposite situations. If base located in north latitude, then SCSS should begin illuminate when sunlight starts fall on sail surface from south. And opposite situation is for bases, which are located in the south latitude. From this, it follows that the solar sail can provide additional light for same area only during the half of Martian year. Inactive time could be used for phase adjustment of the spacecraft position in ASO, since pseudo cylindrical orbit doesn't provide fully stationary position due to variations in altitude (Fig. 8).

The Mars base with coordinate 8° of the north latitude and 46° of the east longitude was investigated. Characteristic acceleration generated by SCSS is $a_0 = 8$ mms$^{-2}$.

Regardless of the variations, average altitude of the spacecraft continuously increases as it's shown in Fig. 9. Alteration of this average value in the first place is connected with axial tilt of Mars, because of it the Sun descends relative to equatorial plane. Thus, the vertical component of the acceleration increases (Fig. 10), meanwhile the parallel component, which is shown in Fig. 11, decreases. During all the time until the winter solstice average value of angle $\theta$ increases as well (Fig. 12), that gives additional expansion of the acceleration produced from solar radiation pressure.

Four plasma propulsion engines SPT-290 and control law for equalization allow us to greatly reduce the divergence in positions of the solar sail and illuminated area. However, this difference in phase angle are unavoidable. At the end of the flight the angle between Mars base and the spacecraft position in equatorial plane is equal $\delta = 25°$. Behavior of this divergence is shown in Fig. 13

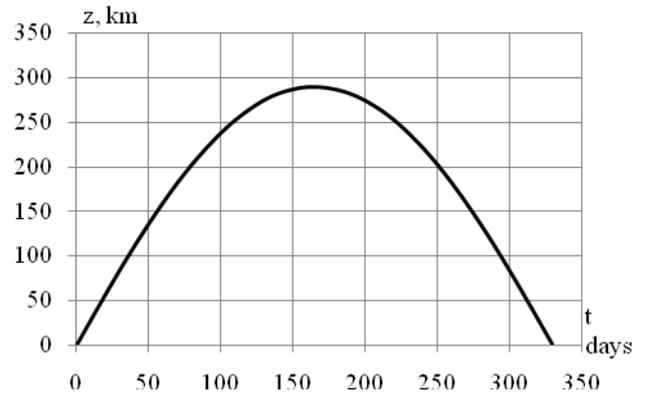

Fig. 9. Average altitude of the SCSS.

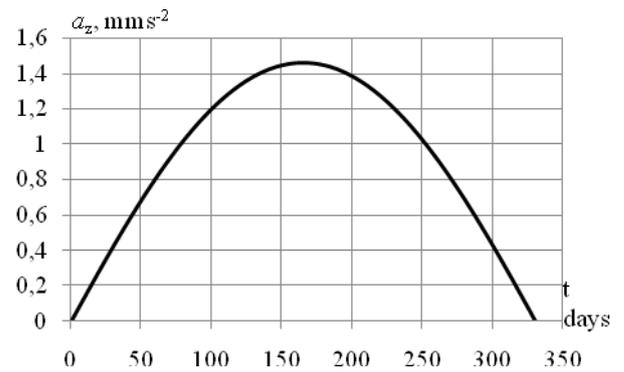

Fig. 10. Average value of the component of the sail acceleration $a_z$ perpendicular to the equatorial plane.

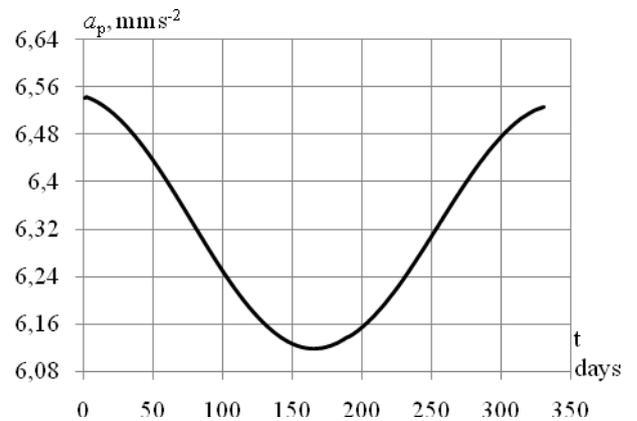

Fig. 11. Average value of the component of the sail acceleration $a_p$ parallel to the equatorial plane.

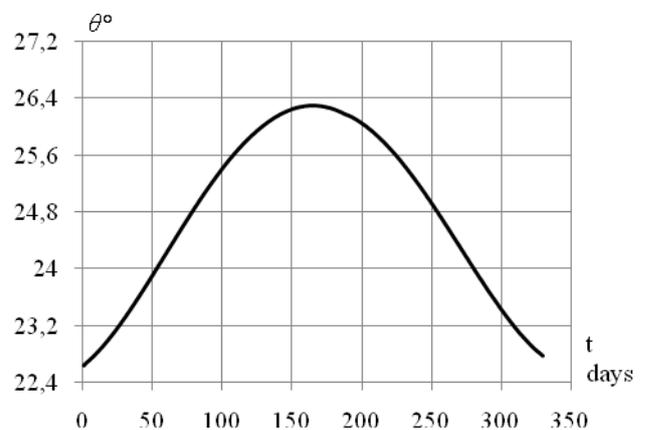

Fig. 12. Average value of the angle $\theta$.

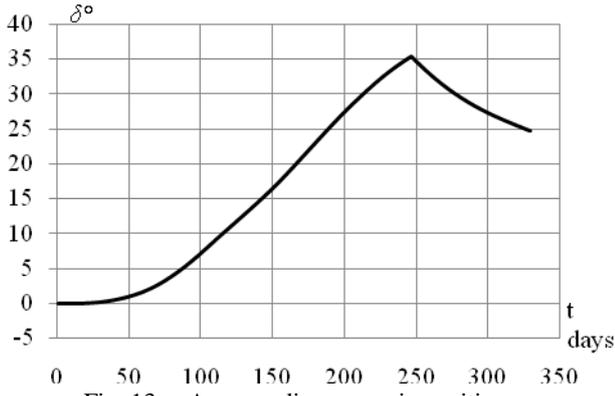
Fig. 13. Average divergence in positions.

### 4.2. Diurnal motion

Change of the vertical component of sail acceleration essentially influences on spacecraft's behavior during daily motion. The more value of the acceleration, the faster it changes due to high variations of the angle $\theta$. But it's necessary to pay attention on initial positions of the SCSS and the Sun, before reflecting sunlight.

Consider case, when spacecraft located in ASO, the Sun is almost at the winter solstice and the solar sail begins reflecting sunlight on the Mars base. The solar sail acceleration creates high amplitudes of altitude variations (Fig. 14). However, the spacecraft doesn't descend lower then equatorial plane, therefore the tilt angle of the orbit is appropriate low. And still, regardless of the amplitude reducing and further stabilization the disturbance, created by these variations, increases divergence in positions of illuminated area and SCSS.

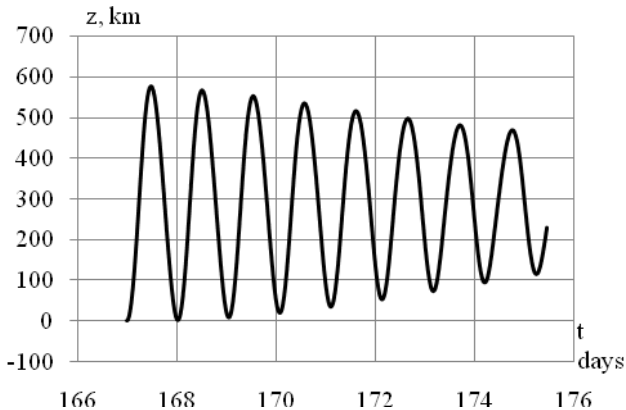
Fig. 14. Rise of the solar sail from areostationary orbit.

Authors suggest to rise the spacecraft beforehand on the required altitude for the stable motion at cylindrical orbit. This altitude can be calculated from Eq. (4) with condition of vertical components equality for both gravitational and solar radiation pressure accelerations

$$a_0 \cos^2(\theta) \cdot n_z^0 = |\mathbf{a}_g| \cdot \frac{z}{|\mathbf{r}|}. \qquad (6)$$

Usually for this equation the angle $\theta$ is defined as average around 24.5°. Direction cosine toward z axis is given from condition of continuously reflection of the sunlight off the solar sail on the Mars base, which is given by Eq. (3).

For previous scenario required altitude is equal $z = 285$ km. Behavior of the SCSS at the same period of time, but with the initial altitude is shown if Fig. 15. Compared with the zero altitude start, the amplitude of the altitude variations is significantly lower. Therefore, it's better to rise spacecraft to the proper altitude before starting reflection of the sunlight on the Mars base. However, when the Sun is located in equatorial plane, it is better to start illuminating from the ASO plane.

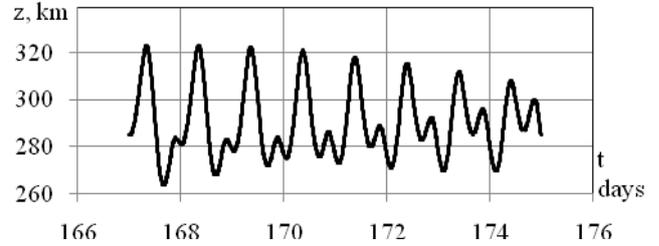
Fig. 15. Keeping the spacecraft on pseudo cylindrical orbit.

### 7. Conclusion

Authors successfully performed simulation of the solar sail motion in condition of reflection the sunlight on established area of the Mars surface and accomplished several researches of SCSS behavior. Motion simulation allows us to make conclusion, that changeable value of the solar sail thrust and its projection on the equatorial plane produce essential disturbances in spacecraft motion and stability of the stationary orbit. In this case, we recommend to use the "space mirror", with a high value of the areal mass to support the certain climate on a Mars base. Unlike the solar sail, the space mirror won't produce such big amount of disturbances, thus it will be easier to keep spacecraft on cylindrical stationary orbit by means of electrically powered spacecraft propulsion.